\documentclass{article}

\PassOptionsToPackage{sort&compress, numbers}{natbib}
\usepackage[preprint]{neurips_2023}
\usepackage{multirow}

\usepackage[utf8]{inputenc} 
\usepackage[T1]{fontenc}    
\usepackage{hyperref}       
\usepackage{url}            
\usepackage{booktabs}       
\usepackage{amsfonts}       
\usepackage{nicefrac}       
\usepackage{microtype}      
\usepackage{xcolor}         
\usepackage{algorithm}
\usepackage{algorithmic}
\usepackage{graphicx}

\title{Improving few-shot learning-based protein engineering with evolutionary sampling}

\author{%
  M. Zaki Jawaid*\\
  EpiCRISPR Biotechnologies \\
  \texttt{zaki.jawaid@epic-bio.com} \\
  \And
  Robin W. Yeo\thanks{ZJ and RY contributed equally to this work.}\\
  EpiCRISPR Biotechnologies \\
  \texttt{robin.yeo@epic-bio.com } \\
  \AND
  Aayushma Gautam \\
  EpiCRISPR Biotechnologies \\
  \texttt{aayushma.gautam@epic-bio.com} \\
  \And
  T. Blair Gainous\\
  EpiCRISPR Biotechnologies \\
  \texttt{blair.gainous@epic-bio.com} \\
  \And
  Daniel O. Hart \\
  EpiCRISPR Biotechnologies \\
  \texttt{daniel.hart@epic-bio.com} \\
  \And
  Timothy P. Daley \thanks{Corresponding Author} \\
  EpiCRISPR Biotechnologies \\
  \texttt{tim.daley@epic-bio.com} \\
}

\begin{document}

\maketitle

\begin{abstract} 
Designing novel functional proteins remains a slow and expensive process due to a variety of protein engineering challenges; in particular, the number of protein variants that can be experimentally tested in a given assay pales in comparison to the vastness of the overall sequence space, resulting in low hit rates and expensive wet lab testing cycles. In this paper, we propose a few-shot learning approach to novel protein design that aims to accelerate the expensive wet lab testing cycle and is capable of leveraging a training dataset that is both small and skewed ($\approx 10^5$ datapoints, $<1\%$ positive hits). Our approach is composed of two parts:  a semi-supervised transfer learning approach to generate a discrete fitness landscape for a desired protein function and a novel evolutionary Monte Carlo Markov Chain sampling algorithm to more efficiently explore the fitness landscape. We demonstrate the performance of our approach by experimentally screening predicted high fitness gene activators, resulting in a dramatically improved hit rate compared to existing methods. Our method can be easily adapted to other protein engineering and design problems, particularly where the cost associated with obtaining labeled data is significantly high. We have provided open source code for our method at \url{https://github.com/SuperSecretBioTech/evolutionary_monte_carlo_search}.

\end{abstract}

\section{Introduction}

The design and optimization of proteins with specific functionality is a long-sought pursuit in protein engineering. Since proteins are composed of sequences of amino acids which ultimately dictate their structure and function, the protein engineering problem can be reformulated as finding the optimal mapping from amino acid sequence $s$ of length $L$ to biological function $f: s \to f(s)$, where we call  $f$ the fitness function.  Finding the optimum of $f$ can be seen as a high-dimensional discrete combinatorial optimization problem \cite{weinreich2005perspective}. 
The enormous size of the protein sequence space (e.g. $20^L$ possible peptides of length $L$; $3.87e110$ for $L = 85$) and the presence of sensitive and sporadic high fitness regions in the fitness landscape \cite{ren2022proximal} makes novel protein design extremely challenging.

The traditional experimental approach involves high-throughput, iterative laboratory methods such as directed evolution \cite{DirectedEvolution_Arnold1998, DirectedEvolutionMethods_Packer2015}, deep mutational scans \cite{DeepMutationalScan_Fowler2014}, and semi-rational design \cite{SemiRational_Chica2005}. However, these methods typically require multiple rounds of engineering and analysis, making them tedious, expensive, and time-consuming \cite{DiversityProblemDirectedEvolution_Wong2006}. Furthermore, the number of variants capable of being tested in even the most advanced laboratories ($\approx 10^5$ to $10^6$) is miniscule in comparison to the size of the total sequence space ; additionally, high-throughput screening can be challenging to implement for some classes of proteins \cite{LowN_Biswas2021}.

In the past decade, the application of machine learning methods to protein engineering problems has been massively successful \cite{MLforProteinEngineering_Xu2020}.
In this context, machine learning models are trained to learn the sequence-to-function map (also called the fitness function) and then used to propose new sequences that maximize the fitness (thus maximizing predicted function). Typically these are two distinct steps, where the fitness is estimated with a machine learning model and then this sequence-to-function map is used to explore the fitness landscape with methods such as Metropolis-Hastings Monte Carlo Search~\cite{LowN_Biswas2021}.

In recent years, other methods such as generative models have been proposed to tackle this problem, including deep generative networks \cite{GenerativeModels_Giessel2022, GenerativeModels_NeuripsPossu2018, VariationalAE_Hawkins2021}, generative adversarial networks \cite{GAN_Repecka2021, GAN_Rossetto2020} and diffusion models~\cite{watson2022broadly}.  In these cases the exploration problem is trivial, as the model produces an embedding in a real, typically low-dimensional, space where sampling from that space is computationally inexpensive. However, generative approaches typically require huge amounts of training data and a large number of positive examples to ensure that the model embeddings are meaningful and so that they do not simply memorize positive examples, an issue that has been widely observed to happen in image GANs~\cite{meehan2020non, gulrajani2020towards}. Given the relatively small number of sequences in our training data and the extreme paucity of positive examples, we anticipated our small and skewed training data would would prove insufficient for a generative modeling approach. On the other hand, transfer learning of large protein language models (LPLMs) has shown success in modeling and designing novel proteins with fitness functions trained on small numbers of positive hits ~\cite{LowN_Biswas2021, madani2023large}. While transfer learning and ML-based sequence-to-function mapping are beginning to receive a good deal of attention, model-guided fitness landscape exploration remains an understudied problem in the context of protein engineering \cite{PrimerSearch_Sinai2020}.

The Metropolis-Hastings Monte Carlo Search (MHMCS) method \cite{MetropolisOG_1953, HastingsOG_1970, PrimerSearch_Sinai2020} is the standard method for the exploration of high-dimensional discrete landscapes, including those generated by machine learning algorithms \cite{LowN_Biswas2021, anishchenko2021novo, castro2022transformer}; however MHMCS suffers from an inability to escape deep local optima. Other approaches for sampling the sequence space include gradient-based sampling \cite{OopsGradient_Maddison2021, ren2022proximal}, and modified Gibbs sampling \cite{SIPF_Sun2022}. While powerful, these approaches require significant computation near the local neighborhood of the fitness landscape and are therefore too computationally intensive for sequences of any significant length, (e.g. gradient-based methods require $19 \cdot L$ computations and Gibbs requires $L$ computations per iteration).

Evolutionary Monte Carlo (EMC) \cite{RealParameterEMC_Liang2001, ProteinFoldingEMC_Liang2001} is an advanced sampling method that draws inspiration from genetic recombination as well as physics-based MCMC techniques. While EMC has previously been used for a variety of sampling tasks \cite{EMCHighFrequency_Spezia2021, EvolutionarySequentialMC_Dufays2016, EMCforClustering_Goswami2007, RealParameterEMC_Liang2001, EMC_Zhang2001}, its potential as an exploratory algorithm for protein design remains unexplored. In this paper, we modify EMC as a search tool for exploring the complex fitness landscape of protein sequences capable of gene regulation, which we call EMC Search (EMCS). While EMCS is much less computationally intensive than gradient-based and Gibbs sampling (and only slightly more intensive than MHMCS), we expect it to benefit from faster convergence (due to parallel tempering) and to provide a more comprehensive and efficient exploration of the fitness landscape (by allowing for interpolation on the molecule level between chains).

Overall, we propose a design strategy for novel protein sequences  using a few-shot transfer learning-based approach. Though our method is generally applicable to a diverse range of problems, we apply our method to the design of small gene activator proteins. 
We previously~\cite{carosso2023discovery} performed a high-throughput screen of protein sequences to discover novel gene activators, and identified less than 200 sequences which validated as positive hits (resulting in a hit rate of $\approx 0.5 \%$). The low number of positive examples presents a particular problem for ML-guided engineering because it is difficult to ensure that the fitness function will extrapolate well outside the small neighborhood of the positive examples in the training set. In this study, we demonstrate that EMCS is not only capable of improving the sequence diversity and novelty of designed sequences, but it dramatically improves the hit rate of the proposed sequences compared to the original high-throughput screen. Additionally, EMCS can be initialized from known hits and still identify candidate sequences that are vastly different than any of the original molecules, while MHMCS has difficulty escaping from the local optima of known hits.

\section{Model and Search}

\begin{figure}[t!]
\centering
\includegraphics[width=\linewidth]{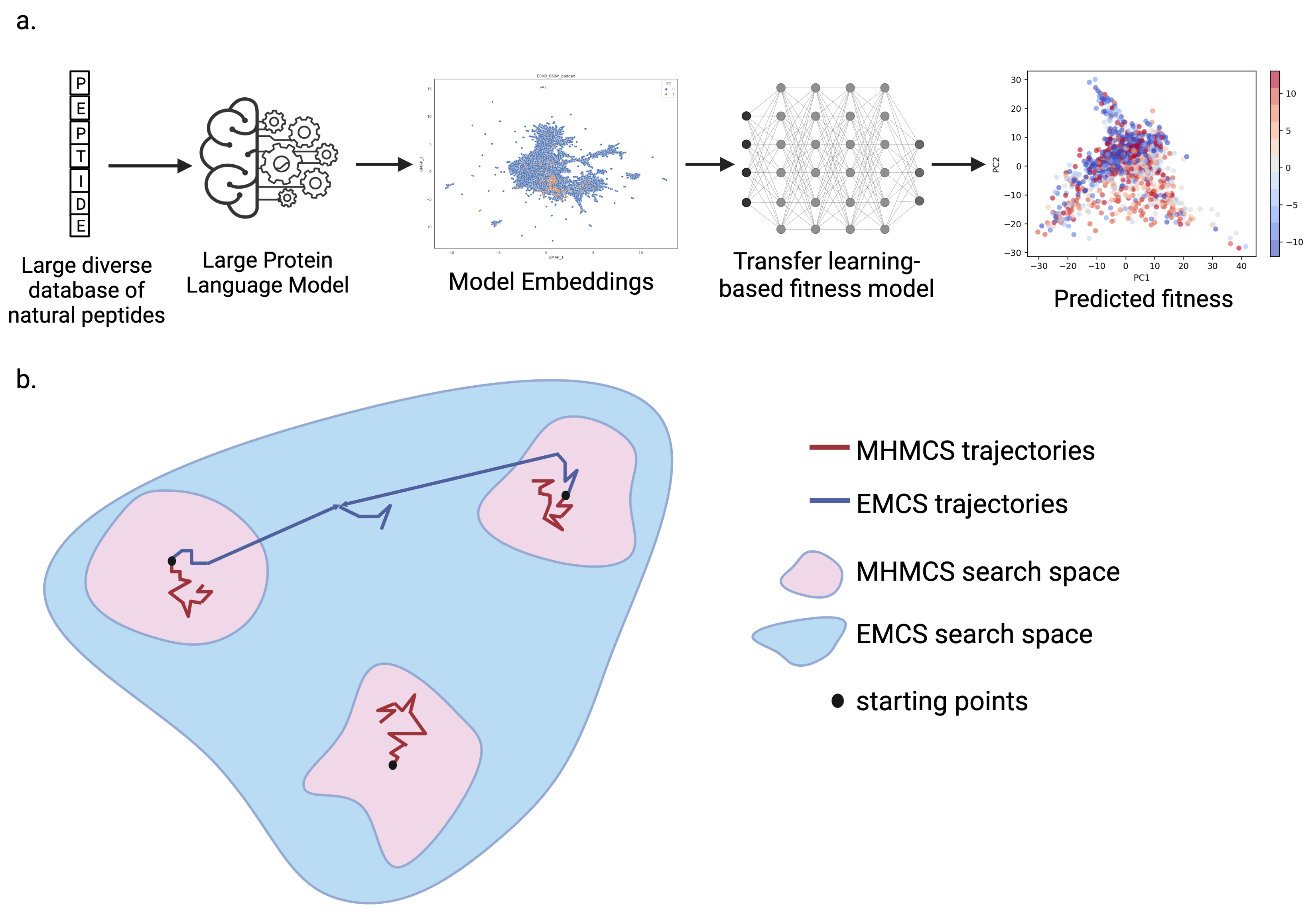}
\caption{\textbf{a.} Transfer learning approach for predicting gene activators from sequence using LPLM embeddings.  \textbf{b.}  Fitness landscape diagram representing the effective search spaces of MHMCS (pink) and EMCS (blue). When initialized at positive hits, MHMCS is constrained to locally search near the starting molecule other high fitness sequences, while EMCS interpolates between multiple starting molecules with varying resolution to optimize the search and escape deep local optima.} 
\label{fig:overview_figure}
\end{figure}

\subsection{Training Transfer Learning-Based Fitness Models}

We previously performed and independently validated a high-throughput screen in which 85 amino acid (85aa) peptides were assayed for their ability to activate a synthetic genetic locus using the dCasMini Gene Expression Modulator System (dCasMini-GEMS) \cite{carosso2023discovery}. 
This resulted in the identification of 173 gene activators ("positive hits") from a training set of 34217 protein sequences ($0.51 \%$ hit rate). Using these data, we sought to train a machine learning model capable of predicting proteins capable of gene activation from sequence alone. Since a data set of peptide sequences is essentially composed of strings of amino acid characters, each peptide sequence needs to be numerically encoded to be used as input to train supervised classification models. We compared OneHot encoding with transfer learning using a 650 million parameter LPLM (ESM-2 model) \cite{Lin2023} as input features for two models: an XGBoost model, where we flatten the features by taking the mean, and a CNN model. In our testing phase we found that transfer learning significantly improved prediction by both models (see Supplementary Tables) and that the sequences proposed by each model appeared to capture different features of our training data.  Indeed, this is not surprising since mean flattening the feature embeddings for XGBoost is equivalent to training on global features of these peptide sequences, while the CNN model is capable of learning local features. We therefore used both models with transfer learning to individually design molecules (Fig~\ref{fig:overview_figure}\textbf{a}), as well as a transfer learning-based ensemble model in order to leverage both the global and local features learned by the XGBoost and CNN models respectively. Specific model architectures and training details are available in the Supplementary.

\subsection{Metropolis-Hasting Monte Carlo Search (MHMCS)}

The MHMCS algorithm operates by proposing a low number of mutations to modify the current molecule and then evaluating the new molecule's fitness; if fitness improves, the proposal is accepted, while, if fitness decreases, the proposal is accepted with probability weighted by the ratio of the proposed fitness to the current fitness.  The latter possibility ensures that sub-optimal moves can be made to ensure that the search is capable of escaping from a local optima, although MHMCS tends to struggle with extremely deep optima.

\begin{algorithm}[t!]

    \caption{Evolutionary Monte Carlo Search (EMCS)}
    \label{EMCS}
    \begin{algorithmic}
    
    \STATE  Select $N$ chains of amino acid sequences $[0, 1, .. ,i, .. , N]$, with corresponding temperature ladder $[T_1, T_2, .. , T_i, .. , T_N]$ such that 
                $T_i \geq T_j$ for $i > j$, with fitness $f(i)$
    \STATE  Set crossover rate $\gamma$ such that $\gamma \subseteq [0, 1)$, and define maximum mutation, crossover, and swap events as $\mu, \alpha, \beta$ 
    \STATE  Set minimum and maximum number of iterations $k_{min}$ and $k_{max}$, respectively
    \STATE  Set convergence condition $f(i) \geq C$ where $f(i)$ is the fitness of sequence $i$
        
    \REPEAT
        \STATE Sample random number $p$ from uniform distribution $[0, 1)$
        \IF {$p > \gamma$}
            \STATE Make random point mutations at $q$ loci for each sequence $i$ to yield new set of proposed sequences denoted by $j$, where $q \in \{1, \ldots, \mu \}$ is chosen uniformly at random
            \STATE Update each sequence by accepting or rejecting each proposed sequence using the metropolis hasting criterion i.e. with probability $min(1,r_{mh})$, where $r_{mh} = \exp (\frac {f(j) - f(i)}{T})$     
        \ELSE
            \FOR{number of crossover events $\alpha$}
                 \STATE Let $i_1, j_1$ be two random sequences corresponding to temperatures $T_i, T_j$. Pick a random crossover locus between [2, N-1], where N is the length of the peptide. 
                 \STATE Propose a set of two sequences $i_{2}$ and $j_{2}$ by crossing over $i_1, j_1$ at the chosen crossover locus. This results in $i_2$ being identical to sequence $i_{1}$ prior to our crossover locus, and identical to sequence $j_{1}$ post our crossover locus. Similarly, $j_{2}$ is identical to sequence $j_{1}$ prior to the crossover locus, and identical to sequence $i_{1}$ post the crossover locus.
                 \STATE For two temperatures $T_i$ and $T_j$ such that $T_{i} \leq T_{j}$, order $i_2$, $j_2$, and $i_1$, $j_1$ such that $f(i_{1}) > f(j_{1})$ and $f(i_{2}) > f(j_{2})$
                 \STATE Accept the new set of sequences $i_2, j_2$ with probability $min(1, r_c)$ where $r_c$ is defined as $r_{c} = \exp \left(\frac{f(i_{2}) - f(i_{1})}{T_{i}} - \frac{f(j_{2}) - f(j_{1})}{T_{j}} \right)$. If accepted, assign $i_2, j_2$ to chains at temperatures $T_i, T_j$ respectively. 
            \ENDFOR
        \ENDIF
                \FOR{number of swap events $\beta$}
                    \STATE Select two sequences $i$ and $j$ at chains corresponding to $T_i, T_j$, such that $j = i \pm 1$, and swap their sequence positions such that $i \to j$ and $j \to i$ with probability $min(1, r_{re})$, where $r_{re}$ is defined as $r_{re} = \exp\left(-(f(i) - f(j)) \left(\frac{1}{T_{i}} - \frac{1}{T_{j}} \right) \right)$
                \ENDFOR
            
    \UNTIL{$iterations > k_{max}$ \OR ($(f_i \geq  C)$ for any sequence \AND $iterations > k_{min})$}
    \end{algorithmic}
\end{algorithm}

\subsection{Evolutionary Monte Carlo Search (EMCS)} 

Evolutionary Monte Carlo Search (EMCS) extends traditional Metropolis-Hastings Monte Carlo Search (MHMCS) by introducing genetic crossover events in a parallel tempering setup \cite{liang2000evolutionary, ProteinFoldingEMC_Liang2001}. In parallel tempering, multiple MHMCS chains are run simultaneously at different temperatures (referred to as a temperature ladder) and are swapped at two randomly chosen temperatures after a predetermined number of iterations. The primary advantage of parallel tempering is that it allows MHMCS to occur over a larger search radius without sacrificing resolution. EMCS builds upon parallel tempering by adding genetic crossover events (domain swapping through chain interpolation). This allows for an even larger search radius (Fig~\ref{fig:overview_figure}\textbf{b}), while also adding the possibility of aggregation of favorable protein domains, which we hypothesize is critical to exploit the small number of positive hits in our training data.

Algorithm \ref{EMCS} details our implementation of EMC as a search tool. EMCS is highly versatile and allows for vastly different exploratory behaviors compared to traditional sampling techniques due to the implementation of a custom temperature ladder, as well as predefined crossover, mutation, and swap rates \cite{liang2000evolutionary}. These parameters can be tuned for more efficient exploration depending on the specific design problem and the complexity of the discrete high-dimensional fitness landscape. 

Each primary iteration in EMCS can potentially change the state of the algorithm in one of three ways, namely, point mutations, swaps, and crossovers between different temperature chains. The possibility of the acceptance of sub-optimal moves for each of these three classes depends on how we define the acceptance criterion. We use $r_{mh}$, the standard Boltzmann Metropolis-Hastings acceptance criterion, for mutation-based moves, which as described earlier, accepts sub-optimal moves with probability weighted by the ratio of the proposed fitness to the current fitness. For swaps between two consecutive chains, we use $r_{re}$, the standard parallel tempering criterion also used in \cite{liang2000evolutionary}. Using this criterion, any proposed swap in which the higher fitness sequence in proposed to move to the lower temperature chain is accepted. In a swap in which a higher fitness sequence is proposed to move to the higher temperature, the move is accepted with probability inversely proportional to the magnitude of the difference of the temperatures of the two chains, as well as the fitness of the two sequences. Finally, the crossover criterion $r_{c}$, also adapted from \cite{liang2000evolutionary}, accepts crossover moves taking into account the difference in fitness between the set of old and new sequences, in addition to the difference of temperatures of the two chains involved in the crossover. For simplicity, we have summarized the behaviour of the crossover criterion in the supplementary, and we note that in general, the crossover criterion penalizes an overall decrease in fitness when taking into account both chains. 

\section{Results}

The protein fitness landscape is known to be highly sensitive, multi-peaked, and rugged \cite{weinreich2005perspective, ren2022proximal}, reflecting the possibility that a complete loss of function can arise due to a relatively small number of point mutations (e.g. mutations in catalytic domains, mutations that cause misfolding, ...). The complexity of this space presents obvious challenges for efficient exploration. Here we compare how EMCS and MHMCS respectively explore the discrete fitness landscape of 85aa proteins capable of gene activation, and evaluate prediction success rates, sequence diversity, and convergence speeds.

\subsection{Experimental Screening}

\begin{figure}[b!]
\centering
\includegraphics[width=0.75\linewidth]{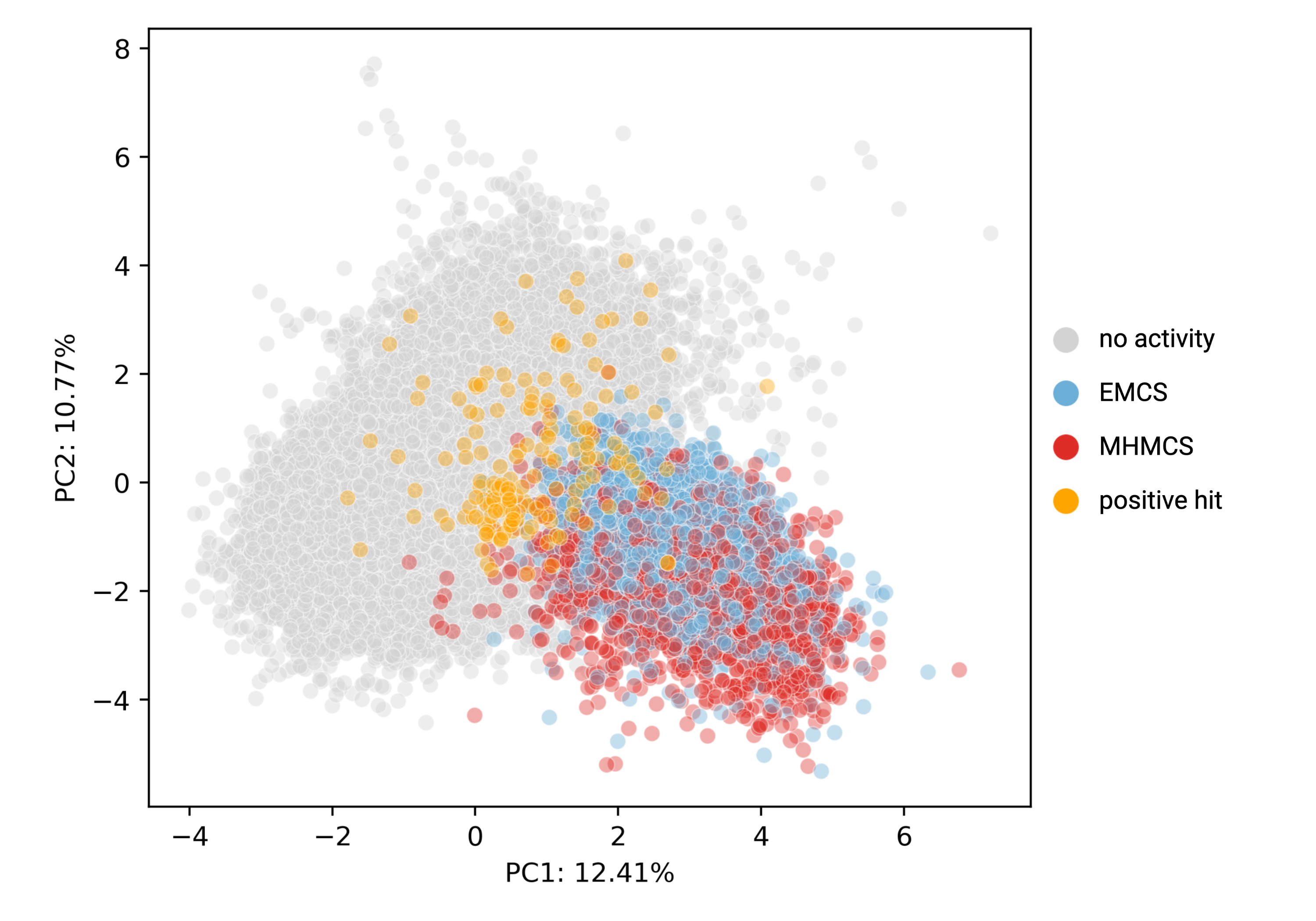}
\caption{\textbf{} Principal Component Analysis (PCA) of original training set (grey and orange points) with novel sequences designed by EMCS (blue) and MHMCS (red) using OneHot encoding.} 
\label{fig:PCA}
\end{figure}

For experimental validation, we used EMCS and MHMCS to design novel proteins using all three of our models (XGBoost, CNN, ensemble). Together, EMCS and MHMCS designed 4600 novel sequences that are largely distinct from the sequence space occupied by the original training data (Fig~\ref{fig:PCA}), confirming that both model-guided sampling techniques are capable of proposing diverse novel proteins. To ensure that we could accurately identify gene activators in our experimental validation, we also included 300 previously validated negative controls (random sequences) to the library. We then experimentally assayed the peptides for their ability to activate a genetic locus (full details of experimental design can be found in Supplementary). In total, we identified 357 positive hits ($7.59 \%$ hit rate) where a positive hit indicates that the peptide was found to activate a synthetic gene reporter significantly over background fluorescence. In contrast, the initial screen had a hit rate of only $0.51 \%$.  If we use the latter number as a proxy for the fraction of naturally occurring 85aa peptide sequences that are capable of gene activation, then our approach increased the baseline hit rate by $\approx 15$-fold. In fact, the best model-guided sampling technique (ensemble model + EMCS from known hits), increased the hit rate $\approx 45$-fold (Table~\ref{table:ensembl_res}) by this metric. Even with initialization from known positive hits, the sequences proposed by EMCS were highly dissimilar from anything in the training set, which suggests that EMCS is capable of escaping deep local optima to efficiently traverse the fitness landscape and identify diverse high fitness peptides. 

\begin{table}[t!]
\begin{center}
\begin{tabular}{|c|c|c|c|c|}
 \hline
Search Method & Initialization & \# Sequences &  \# Positive Hits & Hit Percentage\\
 \hline
 EMCS  & known & 410 & 94 & 22.9\% \\
 EMCS & random & 390 & 39 & 10\% \\
 MHMCS  & random & 200 & 2 & 1 \% \\
 \hline
\end{tabular}
\caption{\textbf{} Positive hit results for the ensemble model}
\label{table:ensembl_res}
\end{center}
\end{table}

\subsection{Sequence Diversity}

To compare sequence proposals between EMCS and MHMCS, we performed an \emph{in silico} sampling experiment where we explore the fitness landscape 4000 times with each algorithm using identical and controlled initial conditions (see Supplementary for additional details).

\begin{figure}[t!]
\centering
\includegraphics[width=\linewidth]{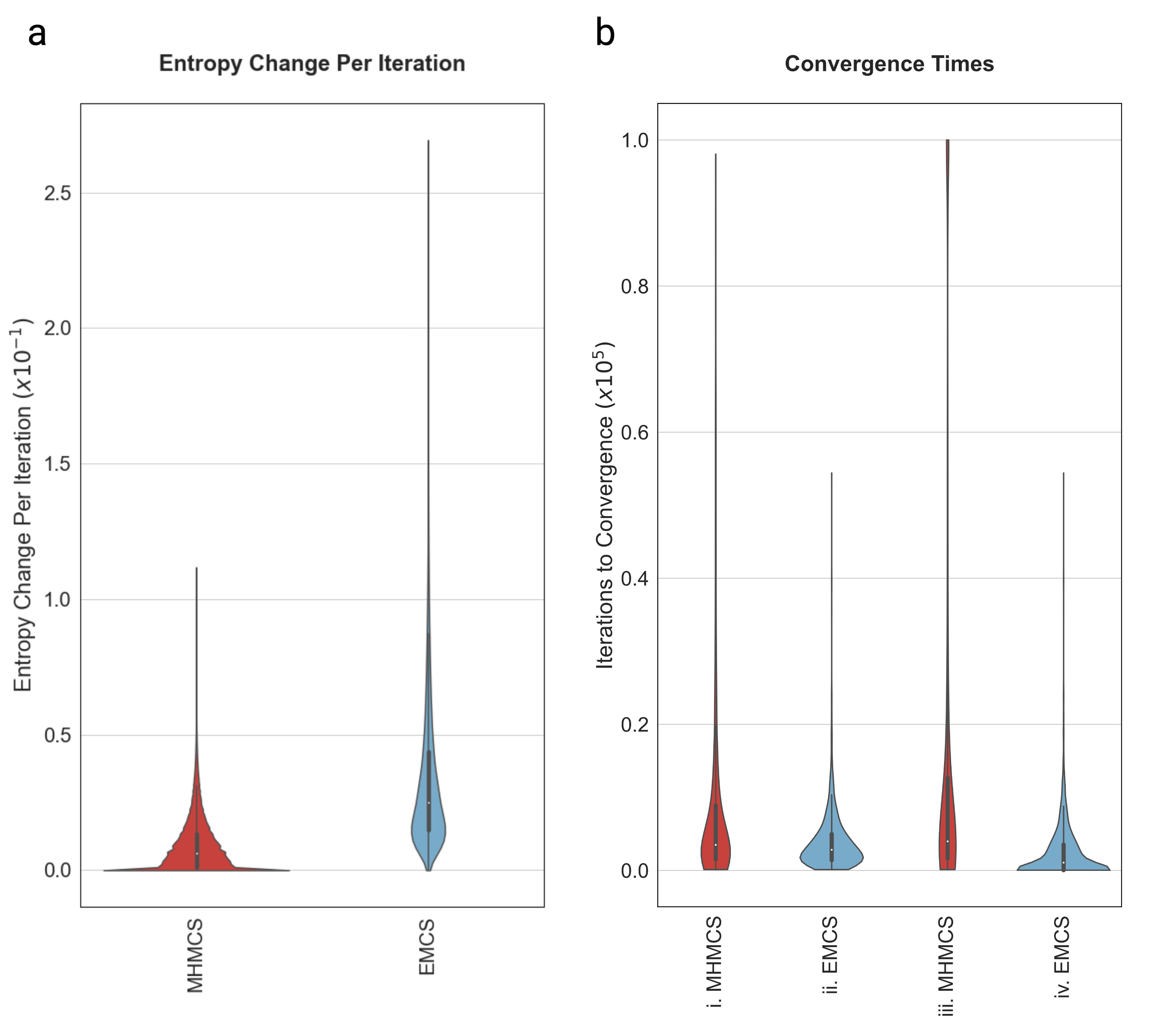}
\caption{\textbf{a.} Entropy change distribution for $10^7$ MHMCS and EMCS primary iterations using default parameters. From an information perspective, EMCS explores a larger region of the fitness space per iteration compared to MHMCS. \textbf{b.} i) Iterations to convergence ($f \geq 0.95$) starting from random pre-defined sequences for 2361 sequences obtained by 2571 MHMCS runs at $T = 2.5 \times 10^{-3}, 10^{-4}$. ii) Iterations to convergence for 1171 sequences obtained by 1171 EMCS runs under default parameters. iii) Number of iterations to arrive at convergence and/or positive hits ($f \geq 0.5$) for 2571 sequences from 2571 MHMCS runs at $T = 2.5 \times 10^{-3}, 10^{-4}$. 210 of the 2571 sequences failed to reach convergence (but succeeded in yielding positive hits ($f > 0.5$)). iv) Iterations to arrive at convergence and/or positive hits for 2720 sequences obtained by 1171 EMCS runs under default parameters, yielding an average of 2.32 positive hits per EMCS run of 4 chains.} 

\label{fig: sequence_figure}
\end{figure}

A unique advantage of EMCS is its ability to identify novel high fitness sequences even when initialized from sequences that were known positive hits (and thus already in a high fitness neighborhood). When initialized from known positive hits, the final edit distances of sequences discovered by EMCS are significantly higher when compared to the sequences discovered by MHMCS using a similar temperature regime (see Supplementary). Consistently, using entropy as a measure of information change, we calculated the average entropy change per iteration of EMCS and MHMCS over $10^7$ iterations (Fig.\ref{fig: sequence_figure}a) and we show that the average entropy change per iteration in EMCS is $\approx 3$-fold higher (using the default parameters of crossover rate of 0.5 and a total of 4 chains) than that of MHMCS (assuming the same mutation rate).

We postulate that the increased proposed sequence diversity and increased entropy per iteration seen with EMCS is due to the genetic crossover steps, where functionally beneficial protein domains can be exchanged between known sequences which are then further refined via point mutations. Escape from local minima is further encouraged by the incorporation of a temperature ladder, which allows for an increase in the search radius. In contrast, MHMCS is restricted to a single temperature and can only access domains in the fitness function that are accessible via point mutations alone. This hinders the ability of MHMCS to converge at a domain that corresponds to a diverse sequence when starting from a known positive sequence because it will require many sub-optimal moves to escape for the local optima of the initial sequence.

\subsection{Convergence}

When initialized at random sequences, EMCS converges 1.25 - 5x faster than MHMCS (depending on choice of temperature and crossover rates, as shown in Fig.\ref{fig: sequence_figure}\textbf{b}) likely due to the algorithm's increased versatility over MHMCS. With default parameters, we achieved convergence for 1171 EMCS runs where we obtained at least one sequence per run that had a fitness $\geq 0.95$. In addition, due to the inclusion of 4 chains, EMCS yielded an average of 2.322 sequences per run that had fitness $\geq 0.5$, thereby giving us a total of N = 2720 sequences of fitness $\geq 0.5$ for 1171 runs. For MHMCS, chains that started at temperatures greater than $2.5x 10^{-2}$ had a minimum failure rate of 50\%, and were dropped from the experiment. When excluding those sequences, we obtained a total of N = 2571 sequences from 2571 runs. 2361 of those sequences had fitness $\geq 0.95$. The remaining 210 failed to reach convergence, but still had final fitness $\geq 0.5$.

\section{Discussion}

In this work, we propose a two step machine learning and sampling approach for protein engineering problems where training data is limited and positive hits are rare. Our method involves leveraging Large Protein Language Models (LPLMs) with transfer learning to estimate a fitness landscape, and then efficiently sampling the fitness landscape with Evolutionary Monte Carlo Search (EMCS) to propose novel high fitness protein sequences. As a proof-of-concept, we apply this approach to the problem of designing small gene activators and demonstrate that our method is capable of successfully designing novel and diverse protein sequences with dramatically higher experimental validation rates when compared to a more traditional sampling method (MHMCS) or baseline discovery from high-throughput screening.

The importance of this approach is magnified when taking into account the complexities of the wet lab testing cycle: a single round of screening involves library design, DNA synthesis, plasmid cloning, viral packaging, cell line infection, fluorescence-activated cell sorting (FACS), DNA library preparation, next-generation DNA sequencing, and downstream bioinformatic analysis. Furthermore, in the field of rational protein engineering, multiple rounds of iterative screening are generally required to discover and validate novel proteins with desired functionality. Given the financial, temporal, and technical costs associated with the wet lab testing cycle, there is obvious value in accelerating this process to reduce the experimental burden of protein engineering. 

The immensity of the protein sequence space coupled with the computational cost of embedding a protein using LPLMs like ESM-2 called for an efficient sampling algorithm that could escape local optima without compromising resolution. The EMC algorithm is ideally suited to this use case, as the incorporation of a temperature ladder allows for the simultaneous existence of multiple acceptance ratios. Furthermore, the genetic crossover steps allow for more efficient exploration of the fitness landscape, as shown by sequence diversity and  average entropy change per iteration of MHMCS vs. EMCS.

We believe that the power of our approach lies in the combination of transfer learning via LPLMs and EMCS. Since LPLMs are trained on an immense number of diverse protein sequences, modern LPLM embeddings implicitly contain a wealth of features describing a protein's biochemical, biophysical, evolutionary, and even 3-dimensional structure information \cite{Lin2023}; as such, we reason that LPLM embeddings of novel proposed sequences are capable of capturing the predicted functional consequences of genetic crossovers from EMCS such that swaps resulting in misfolded or non-active proteins are assigned low fitness and thus not selected by EMCS. Conversely, potential swaps and domains that can act synergistically will be assigned a high fitness by our semi-supervised transfer learning-based model and selected for by EMCS, even if those domains are not evolutionarily related. In contrast, since GANs and diffusion models sample from a low-dimensional latent space, and then pass the sample through the model to obtain the proposed sequence, only sequences that are close to the training data in latent space can be designed by these methods; additionally, there's no guarantee that high synergy domains will be close in the latent space (especially if they're not evolutionarily related) limiting the potential diversity of sequences that can be proposed by generative algorithms trained on limited and skewed training data.

We believe our framework has a number of advantages over both prior ML-guided protein design approaches with traditional sampling techniques as well as the classic laboratory protein engineering approach. Firstly, assays that screen diverse, natural proteins for peptides of specific function typically have extremely low hit rates whereas novel sequences proposed by our approach had significantly higher hit rates in the validation experiment. Additionally, the small number of positive hits in the training data of protein engineering problems inherently limits the accuracy and generalizability of the fitness function; by leveraging information from LPLMs and incorporating multiple positive hits in the proposal of novel sequences through EMCS domain swapping, we believe our approach is capable of attenuating these disadvantages. Finally, though our proof-of-concept involved the design of relatively small proteins, we anticipate that our approach will generalize especially well to protein engineering problems involving larger proteins with multiple well characterized domains. While we aim to extend our approach to the application of larger proteins, our sampling algorithm will first need to be modified and optimized as random swaps within larger proteins are increasingly likely to result in low fitness predictions due to the presence of longer conserved domains. The approach described here should be of benefit to the wider scientific community, especially those involved in protein engineering challenges, and has the potential to accelerate the design and testing of novel proteins for a variety of purposes including therapeutic medicines.

\section{Acknowledgements}

The authors would like to thank Chris Still and Spencer Lopp (Biological Data Science, EpiCRISPR Biotechnologies) for assistance with code review and data analysis pipelines, as well as Xiao Yang and Giovanni Carosso (Technology Development, EpiCRISPR Biotechnologies) for useful conversations. The authors would also like to acknowledge Gladstone Genomics Core for next generation library sequencing.  

\section{Competing Interests Statement}
The authors are affiliated with EpiCRISPR Biotechnologies as employees and hold equity in the company. Several authors are inventors on provisional patent applications for the small gene activators described in this work.  

\newpage

{
\small

}
\end{document}